**The Neoplasia as embryological phenomenon and its implication in the animal evolution and the origin of cancer. I. A presentation of the neoplastic process and its connection with cell fusion and germline formation**

Jaime Cofre[1,*], Kay Saalfeld[2]

[1] Laboratório de Embriologia Molecular e Câncer, Federal University of Santa Catarina, room 313b, Florianópolis, SC, 88040-900, Brazil

[2] Laboratório de Filogenia Animal, Federal University of Santa Catarina, room 103b, Florianópolis, SC, 88040-900, Brazil

[*] Corresponding author. Laboratório de Embriologia Molecular e Câncer, Universidade Federal de Santa Catarina, Sala 313b, Florianópolis, SC, 88040-900, Brazil

E-mail addresses: jaime.cofre@ufsc.br, kay.saalfeld@ufsc.br

**Abstract**

The decisive role of Embryology in understanding the evolution of animal forms is founded and deeply rooted in the history of science. It is recognized that the emergence of multicellularity would not have been possible without the formation of the first embryo. We speculate that biophysical phenomena and the surrounding environment of the Ediacaran ocean were instrumental in co-opting a neoplastic functional module (NFM) within the nucleus of the first zygote. Thus, the neoplastic process, understood here as a biological phenomenon with profound embryologic implications, served as the evolutionary engine that favored the formation of the first embryo and cancerous diseases and allowed to coherently create and recreate body shapes in different animal groups during evolution. In this article, we provide a deep reflection on the Physics of the first embryogenesis and its contribution to the exaptation of additional NFM components, such as the extracellular matrix. Knowledge of NFM components, structure, dynamics, and origin advances our understanding of the numerous possibilities and different innovations that embryos have undergone to create animal forms via Neoplasia during evolutionary radiation. The developmental pathways of Neoplasia have their origins in ctenophores and were consolidated in mammals and other apical groups.

**Introduction**

An aspect undoubtedly overlooked when discussing multicellularity and the emergence of metazoans, mostly as a result of disregard by embryologists themselves, is that animal multicellular organization could not have evolved without the formation of the first embryo. The most basic requirement of metazoan organization is the fusion of an oocyte and a sperm cell to create an embryo. It is well accepted that cell fusion in unicellular holozoans was a turning point in the transition to embryonic multicellularity. In general, cell fusion occurs at important points of transition during animal development (Müller, 1990), and several forms of symbiosis can be interpreted as fusion events leading to biological innovation (Margulis and Fester, 1991). We propose that polyspermy (which is in fact more appropriately referred to as multiflagellate fusion, given that the flagellum of a sperm cell is used for movement and flagella participating in the first fusion event served a sensory function), a phenomenon whose relevance to evolution has been greatly underestimated (Salinas-Saavedra and Vargas, 2011), might have been a major determinant of biological (cellular processes) and physical consequences (Newman, Forgacs and Müller, 2006) resulting in the appearance of the first animal embryo. Multiflagellate fusion, however, is not the focus of this investigation and will be discussed in a separate article.

The first embryo, as a prerequisite, carried in its most basic essence the ability to differentiate both germline components (male and female), which inevitably leads us to define it as a hermaphrodite organism. Hermaphroditism is a consequence of fusion during the first õfertilization.ö Fused cells, on the other hand, are the result of a long experience of cellular differentiation in unicellular organisms having meiosis as part of their life cycle. Some may find it difficult to accept that we can discuss the concept of cell differentiation in unicellular organisms, but the feasibility of this idea becomes pristine clear when we consider, for instance, a unicellular organism that multiplies by meiosis to generate two phenotypically different cells through a process of anisogametogenesis (Tschermak-Woess, 1959) or when we observe multicellular associations formed by several phenotypes throughout their life cycle (Matt and Umen, 2016; Mathavarajah *et al.*, 2021).

The theoretical framework of this article is based on the premise that the formation of the first embryo was not a slow and gradual process, but rather a fast, intense, and consistent process that abruptly generated extremely complex cellular and tissue structures. Some of the most relevant observations include cells that undergo epiboly, collectively coordinated invagination, and epithelialómesenchymal transition (EMT) for the formation of muscle cells and neurons, among others, as well as the establishment of embryonic polarity by the embryo itself during development. Thus, it is possible to affirm that, from a biological and philosophical point of view, the whole process of rapid construction of the embryo is encoded in the biology and physics of initial fusion cells. In this article, we will also underpin the view that cell potency is revealed during morphogenesis and, therefore, differentiation is the result of embryonic architecture and construction.

It is impossible not to marvel at embryogenesis when manipulating animal embryos in the laboratory, observing fertilization taking place, and watching the fusion event disclose the structural coherence of the animal we have just fertilized. This important property of embryogenesis is known as potency or totipotency. Some authors reduce the meaning of totipotency to the ability of a cell to differentiate into all cell types of an organism, but we argue that it is more than that. Totipotency is interconnected with morphogenetic movements that attain maximum expression in the construction of different animal forms.

We further speculate that basic and transcendental elements of cancer (such as Myc, Src, Ras, and Abl orthologs, among others), which were already present in unicellular holozoans (Suga *et al.*, 2014; Sebé-Pedrós, Peña, *et al.*, 2016), participated in the construction of body forms in the first metazoan. It is logical that such elements worked in a distinct, independent manner in unicellular holozoans, serving specific purposes in sporogenesis, gametogenesis, proliferative filopodial stage, or aggregative filopodial stage (Sebé-Pedrós, Peña, *et al.*, 2016); that is, these elements acted in a completely different context than that of cancerous diseases as we understand them today. Nevertheless, a fundamental part of our hypothesis is that the fusion of unicellular holozoan cells allowed the recruitment and co-option of basic elements of cancer to the main core of the first embryo.

From the point of view of embryology and phylogeny, when discussing Neoplasia as an evolutionary engine, it is crucial to focus our attention on six fundamental characteristics of cancer cells. First and foremost, it is important to bear in mind that, in cancer, cell division is out of control, being excessive and continuous (Matthews, Bertoli and de Bruin, 2022). The second important characteristic is the ability of cancer cells to metastasize and expand beyond tissue boundaries. The third characteristic comprises alterations in adhesion properties of the cell membrane that determine the cell's ability to proliferate, migrate, and metastasize. The fourth characteristic refers to altered cell metabolism. The fifth is that cancer has always been considered a disease of cell differentiation (Markert, 1968). Finally, the sixth characteristic is that cancer is associated with mechanosensory systems that transform mechanical stimuli received by the extracellular matrix or cytoskeleton into chemical signalsô a process known as mechanotransduction (Paszek and Weaver, 2004; Paszek *et al.*, 2005; Aguilar-Cuenca, Juanes-García and Vicente-Manzanares, 2014; Fattet *et al.*, 2020).

From a conceptual point of view, we propose recruitment (co-option) of cellular processes that were independent in unicellular organisms; that is, the formation of an embryo as an evolutionary innovation occurred by co-option of different processes at a cellular level (Müller, 1990, 2021). This concept of recruitment or co-option, as we understand it, is well consolidated in the original ideas of Gould and Vrba, being of great relevance to our hypothesis because it differs from adaptation models and is not restricted to phenotypic traits (Gould and Vrba, 1982). We believe that the concept of recruitment or co-option is sufficiently broad to contemplate developmental pathways (Chipman, 2021; Müller, 2021). In our proposal, recurrent transitions from unicellularity to multicellularity in the life cycle of some fungi (Nanjundiah, 2016; Ros-Rocher *et al.*, 2021), as well as their different forms (Margulis *et al.*, 1989), can be regarded as developmental stages and serve as key points for understanding the emergence of the first animal embryo.

Co-option of developmental pathways is a process by which pre-existing characteristics of unicellular Holozoa would effect evolutionary changes, providing an internal cellular environment conducive to a great revolution. Thus, co-option contributes to the consolidation of a module that we will hereafter refer to as neoplastic process (NP) when talking about cell dynamics or neoplastic functional module (NFM) when talking about chromatin structure as influenced by biophysical phenomena. The NP produces multiple effects at the cellular level, including gene regulation, cell behavior, differentiation, and movement, as well as the relative timing of these events, as already predicted by Ariel Chipman (Chipman, 2021) and Gerb B. Müller (Müller, 2021) in studies addressing the exaptation of developmental pathways. Co-option as proposed in our hypothesis is coherent with the three-dimensional (3D) architecture of the nucleus and its influences on gene expression (Cremer and Cremer, 2001; H. Wang *et al.*, 2018; Belyaeva *et al.*, 2022). The presence of topologically associated chromosomal domains constructed during embryonic development (Flavahan, Gaskell and Bernstein, 2017) and deconstructed in cancer (Hnisz *et al.*, 2016) are a fundamental tenet of our theoretical framework (Figure 1). Therefore, our hypothesis argues for a recruitment (i.e., association of chromosomal domains) of processes of proliferation, invasion, adhesion, metabolism, differentiation, and mechanosensing, thereby establishing an NFM with specific 3D chromatin organization. Furthermore, our proposal incorporates the concept of modular exaptation (Andriani and Carignani, 2014) as an innovation mechanism allowing the transformation of this NFM during the first embryogenesis, leading to the appearance of the first embryo as an emergent property (Stephan, 1998).

In unicellular Holozoa, there is evidence of a functional premodule conserved throughout animal evolution that is compatible with our proposal. Chromatin immunoprecipitation sequencing analyses in protists showed Brachyury regulating promoters of actin polymerization and cell motility/amoeboid movement, representing anticipation of functions seen in metazoan gastrulation (Sebé-Pedrós, Ballaré, *et al.*, 2016). Similar conservation was observed in Myc target networks regulating ribosome biogenesis in unicellular Holozoa. A comparison was made between Brachyury targets identified in the life cycle of *Capsaspora owczarzaki* (Sebé-Pedrós, Ballaré, *et al.*, 2016) and Brachyury's known targets in activin A-treated mouse stem cells (Lolas *et al.*, 2014), which serve as an *in vitro* model of primitive streak formation. Of note, this comparison did not account for physical characteristics and morphogenetic deformations not initially present in the cell culture.

Nevertheless, examination of isolated cells revealed a functional module already present in unicellular Holozoa, suggesting the existence of a Brachyury downstream target network involved in cell migration in premetazoan lineages or, in the case of the Myc target network, involved in ribosome biogenesis, which was conserved during evolution. Some factors co-opted into the NFM had different functions throughout evolution, such as the Hippo pathway, which regulates actin cytoskeleton dynamics in *C. owczarzaki* and participates in cell proliferation in metazoans (Phillips *et al.*, 2021). In metazoans, Myc was co-opted to control cell growth and proliferation, among other functions. It is believed that chromatin remodeling is an essential part of the mechanism by which Myc activates or represses its target genes, placing this important regulator at the center of tumorigenesis (Van Riggelen, Yetil and Felsher, 2010). Based on these observations, we speculate that, at the beginning of animal evolution, there was a remodeling of the activities that formed the NFM.

A functional premodule has also been observed in ichthyosporeans. In the cellularization stage, *Sphaeroforma arctica* shows temporal co-expression of proteins associated with the actin cytoskeleton and cell adhesion molecules, such as the integrin receptor and α- and β-catenins (the latter of which is known to participate in cell adhesion together with cadherins). Actomyosin contraction and mechanical strength were also found to play an important role during this stage of transient multicellularity (Dudin *et al.*, 2019). This would represent, in unicellular Holozoa, an anticipation of the role of Physics in the formation of the metazoan embryo. To gain a better insight of the physical aspects of the functional module, it is important to compare the cellularization stage of ichthyosporeans with embryos in epiboly and the contractile ring of zebrafish. Such an investigation is deemed necessary because Brachyury is controlled by β-catenin-dependent external mechanical strain, as reported in *Nematostella vectensis* (Pukhlyakova *et al.*, 2018). Thus, actin is a Brachyury target (Sebé-Pedrós, Ballaré, *et al.*, 2016), but Brachyury responds to a biophysical property that acts on the actin cytoskeleton, creating something of an epigenetic loop with a clearly biophysical component. The properties of the actin cytoskeleton, such as contractility, intercellular contractile ring (Behrndt *et al.*, 2012; Hernández-Vega *et al.*, 2017), electrical birefringence (Kobayashi, Asai and Oosawa, 1964), ion wave propagation along microfilaments (Tuszy ski *et al.*, 2004), and regulation of genome organization (Uhler and Shivashankar, 2017b), are evolutionary spandrels (Gould, 1997) that are not encoded in the genetic material and for which there are no known cis-regulatory elements targeted by transcription factors. Therefore, Physics is assumed to have participated in NFM formation, translating morphogenesis into the cell nucleus and, as we will discuss later, establishing a mechanical memory used to reconstruct the process in the following generation, thereby constituting this emergent property we call the animal embryo.

Another characteristic of the NFM is that it links processes with a high degree of duality. For example, as previously mentioned, molecular pathways involving Myc, Src, Ras, and Abl orthologs were recruited in unicellular Holozoa (Suga *et al.*, 2014; Sebé-Pedrós, Peña, *et al.*, 2016) and probably induced high rates of cell proliferation and expansion. There is also strong evidence supporting the appearance of many cancer genes during the early phases of the emergence of metazoans (Domazet-Lo o and Tautz, 2010; Chen *et al.*, 2015), prompting the belief that the predominant phenomenon at the beginning of animal life would be a neoplastic expansive force. However, there was also the recruitment of molecular pathways involving tumor suppressors, such as p53, retinoblastoma protein (pRb) (Bartas *et al.*, 2019), and Hippo (Sebé-Pedrós, Peña, *et al.*, 2016), acting in conjunction as a brake or barrier to the neoplasm.

At this point, it becomes necessary to clarify that the NP, as it is approached in this article, does not involve any genetic mutation. Somatic mutation is a concept deeply ingrained in oncology and entwined in the molecular foundations of cancer (Weinberg, 1983;

Hanahan and Weinberg, 2000). A classic example of somatic mutations, as well as of their pivotal role in oncogenesis, is found in Ras mutations, present in 19% of cancer patients (Prior, Hood and Hartley, 2020). It is also worth mentioning the relevant and well-recognized role of the non-mutated protein (wild-type Ras) in promoting exponential growth in cancer cell models (T24, MIA PaCa-2, and RD cell lines), whose effects within the cell are non-redundant and independent from those of mutated Ras isoforms (Young, Lou and McCormick, 2013). Interestingly, even in the absence of Ras mutants, activation of wild-type Ras can lead to cancer if some of its negative regulators are lost, such as Nf1, Gap, and Spry proteins (Zhou, Der and Cox, 2016). Thus, modulating the oncogenic activity of non-mutated Ras is among the challenges and future strategies for cancer treatment (Sheffels and Kortum, 2021).

Something similar is observed with Src (Frame, 2002) and Abl (Greuber *et al.*, 2013; Luttman *et al.*, 2021). For both proteins, it is an increase in expression or catalytic activity, rather than a mutation, that influences cell growth, adhesion strength, and even metastasis. Therefore, Src or Abl dysregulation in cancer cells may increase tumor growth (Frame, 2002) and/or enhance migratory or invasive potential (Greuber *et al.*, 2013). In solid tumors, activated Abl kinases promote invadopodium formation, invasion, and cellular processes implicated in EMT and metastasis (Luttman *et al.*, 2021). By contrast, under physiological conditions, Abl kinases are surprisingly recognized to participate in the orchestration of epithelial morphogenesis, cell migration, and cytoskeleton rearrangement (Bradley and Koleske, 2009). These observations demonstrate the modular characteristic of the proposed NP, whereby recruited components are always simultaneously involved with proliferation, adhesion, and invasion, and therefore participate strictly in cellular processes, unlike mutated genes. Cancer as a disease emerged temporally after, and as a consequence of, NP participation in the formation of the first embryo. Or, in other words, some genetic mutations are a consequence of cancer, not its origin.

Another facet of the duality of co-opted processes is denoted by the recruitment of developmental pathways associated with disintegrating amoeboid movements resembling "metastasis," as seen in the life cycles of unicellular Holozoa (Sebé-Pedrós *et al.*, 2013; Suga and Ruiz-Trillo, 2015). When co-opted together with cadherins and the integrin adhesome (Hehenberger *et al.*, 2017), which oppose these disintegrating and disruptive growth forces, such pathways would act as a counterbalance, limiting the disorganization of the first embryonic tissues. Thus, the first animal embryo was formed by exaptations within the NFM, with embryo cells "learning" to grow and move together.

The first exaptation occurred in proliferation pathways linked to structural elements such as cadherins. This allowed cells to remain together and provided the conditions for appropriate, controlled growth. Force generation by actomyosin networks and force transmission along adhesion complexes are two self-organized phenomena that drive tissue morphogenesis; therefore, it can be said that mechanical forces locally regulate cell shape (Heisenberg and Bellaïche, 2013). Cell–cell interactions, such as adhesion, tend to increase the cell–cell contact surface by modulating surface tension (Maître *et al.*, 2012). Evidence suggests that the flattening of simple epithelia, such as those that undergo epiboly during embryonic development, is regulated by cell proliferation at a different orientation, which in turn is modulated by geometric constraints produced by mechanical coupling of cells (Xiong *et al.*, 2014). By drawing on these findings, it is possible to imagine, in the context of cells with intrinsic disaggregating potentials growing together, the occurrence of the first collective migratory movements, such as epiboly.

Secondarily and at a later point in time, the extracellular matrix (ECM) and all components necessary for its remodeling were rapidly developed. These structures were co-opted into the NFM in a process that also comprised exaptation (Gould and Vrba, 1982), necessary for morphogenetic movements and the formation of important tissue structures in the embryo. Our hypothesis is supported by the well-known spatial-temporal pattern of ECM assembly, an event coinciding with the initiation of morphogenetic movements. In amphibians, for example, the beginning of fibronectin fibril assembly, in the roof of the blastocele, precedes mesoendodermal migration (Lee, Hynes and Kirschner, 1984; Nakatsuji, Smolira and Wylie, 1985; Winklbauer, 1998), and further remodeling is needed for the migration process to occur (Davidson, Keller and DeSimone, 2004; Rozario *et al.*, 2009). In *Danio rerio*, whose epiboly is much more similar to that of ctenophores (Bruce and Heisenberg, 2020), the formation of a fibronectin and laminin matrix begins at about 65% epiboly, and fibril formation (fibrillogenesis) commences at 85% epiboly, coinciding with the beginning of convergent extension movements (Latimer and Jessen, 2010). Thus, ECM formation and reshaping are extremely dynamic processes, both spatially and temporally (Rozario and DeSimone, 2010; Walma and Yamada, 2020). This exaptation resulted in the emergence of the collective cell movements (Friedl and Gilmour, 2009) of convergent extension, invagination, and EMT, taking place in a controlled manner for the construction of the animal's body.

In other words, after NFM recruitment, two exaptations were required to achieve the necessary balance for maintaining animal organization and not falling into a pattern of unbridled growth or dispersion by amoeboid movements ("metastasis"). The success in controlling this neoplastic force of growth and expansive migration resulted, within the NFM, in the emergence of a higher-level property known as the animal embryo. When equilibrium is reached, there arises, implicitly and masked by embryonic organization, the disease we call cancer. For those who love evolution, cancer represents the unveiling of cellular processes involved in our evolutionary origin, for in it we can perceive the strength of Neoplasia as well as the intense efforts for its containment, generating the emergent properties of the embryo. Embryo and cancer are therefore two sides of the same coin, of the same process. And Neoplasia can be regarded as the evolutionary force at the heart of animal evolution– one that we need to clearly distinguish from a disease. Neoplasia is a free force, and cancer a force contained by animal organization.

If we accept the premises of NP as an evolutionary engine imbued with the essence of animal formation, we can explain the rapid evolution of metazoan forms, the extraordinary chromosomal rearrangement leading to the formation of laminin IV, Wnt, Fgf8, and Notch at the beginning of animal life. We will be able to explain the numerous rearrangements of mitochondrial DNA that took place in the first invertebrates and are detected in cancer patients. We will also be able to resolve the two strong emergence peaks of cancer-related protein domains, the first at the origin of the first cell and the other at the time of appearance of metazoan multicellular organisms (Domazet-Lo–o and Tautz, 2010). Finally, the recognition of cancer as part of the essence of animal life would explain why the immune system cannot efficiently react to the disease. Given that immune cells intrinsically distinguish what is proper and inherent to animal constitution, there would be no tricks of cancer cells to escape the immune system (Galassi *et al.*, 2021; Jalalvand, Darbeheshti and Rezaei, 2021), but rather a continuous acceptance and collaboration to allow the existence of the animal itself and the coherence of the immune system.

When we fertilize an animal oocyte, we can feel the dizzying force of Neoplasia in its incessant drive for growth and organized EMT. If we release the bonds that maintain cells together in equilibrium, we can feel the maelstrom of what we call Neoplasia in its most traditional form, characterized as the cancer disease. In our view, G. Barry Pierce brilliantly expounded how much Embryology is involved in cancer, noting that a teratocarcinoma is able to differentiate and evolve into a benign form (Pierce, Dixon Jr and Verney, 1959). This was against the grain of what everyone wanted to believe, that cancer cells cannot differentiate:

> "I believe that the ultimate cure for cancer will be through the re-regulation of malignant stem cells to benign stem cells using the principles of embryonic induction and the growth factor action that occurs at the time of organogenesis." (Arechaga, 2003)

This is a simple and brilliant way to realize that a force said to be malignant can contribute to a balanced organization coherent with the maintenance of tissue organization. Basically, we are talking about Neoplasia and how this force, when properly conducted, might have originated the most wonderful diversity of biological forms seen in metazoans.

**Conditions for the emergence of the first animal embryo**

The formation of the first embryo required five fundamental conditions: (i) induction of embryogenesis by fusion of phenotypically different cells (later co-opted as fertilization by flagellated gametes), (ii) the possibility of reconstructing the process coherently and systematically over evolutionary time, (iii) emergence along with Neoplasia from the initial cellular structure (polyploid multinucleate syncytial zygote) through co-option of developmental pathways present in unicellular Holozoa, (iv) coupling of cellular systems to break Neoplasia so as to achieve the structural coherence intrinsic to embryo morphogenesis and organogenesis, and (v) environmental, epigenetic, and physical conditions present in the Ediacaran ocean. Each of these five conditions are intertwined with biophysical phenomena, inherent and fundamental to the emergence of animal life.

*The first condition: cell fusion in unicellular Holozoa*

In basal metazoans, embryos only form from oocytes fertilized with spermatozoa. This is a clear manifestation of the link between embryonic formation and sexual differentiation of gametes. This is important to emphasize, given that we are discussing a cellular biological system, that is, zygotes produced by cell fusion, and not simply a genetic system. Parthenogenesis, on the other hand, is an embryological process acquired a little later in animal evolution (Simon *et al.*, 2003; Jarne and Auld, 2006) and based on some ecological requirements (Cuellar, 1977).

It is crucial to search for conditions for the occurrence of meiosis among protists. A vast number of protists are known to undergo sexual reproduction (Grell, 1973; Heywood and Magee, 1976; Raikov, 1995). Meiosis and/or fertilization have been described in the lobose testate amoeba *Arcella*, the filose naked vampyrellid amoeba *Lateromyxa gallica*, chrysophytes, prymnesiophytes, xanthophytes, dinoflagellates, trypanosomatids, piroplasmids, microsporidia, Myxosporidia, and Actinomyxidia, as amply described by Raikov in 1995 (Raikov, 1995). Furthermore, a cryptic sexual cycle and a meiosis-like recombination was inferred in *Leishmania* from population genetic studies and by laboratory crosses (Inbar *et al.*, 2019; Louradour *et al.*, 2022). Of the groups closest to animals, the choanoflagellate *Monosiga brevicollis* has meiotic genes (Carr, Leadbeater and Baldauf, 2010) and the choanoflagellate *Salpingoeca rosetta* has sexual reproduction with meiotic recombination (Levin and King, 2013). The taxonomic names of higher protists are those adopted by Lynn Margulis (Margulis *et al.*, 1989) and thoroughly reviewed by Sina Adl (Adl *et al.*, 2019).

Raikov classifies sexual protists into three categories: (i) haplonts that undergo zygotic meiosis (Figure 2a), (ii) diplonts with gametic meiosis, and (iii) heterophasic forms with intermediate meiosis. Zygotic meiosis occurs during the first divisions of the zygote, and only the zygote, which is often encysted, is diploid. The vegetative stages are haploid and produce (or become) gametes without meiosis. Gametic meiosis occurs during gamete formation; vegetative cells are diploid and only gametes are haploid. Forms with intermediate meiosis show alternation of two generations, a haploid that produces gametes and ends with fertilization (karyogamy) and a diploid that ends with meiosis, restoring haploidy. Both generations are represented by vegetative forms (Raikov, 1995). In any case, meiosis linked to karyogamy occurs in most groups of protists, having been deeply studied. These observations support the hypothesis that sexuality is a basic characteristic of all protists and has been reduced in some specialized groups (Grell, 1973). In most taxa, meiosis falls into a standard basic type, with pre-meiotic DNA synthesis, initial parallel chromosome pairing, crossing over (which is sometimes lost secondarily, as in achiasmatic meiosis), and synaptonemal complex formation (Grell, 1973; Raikov, 1995).

The meiosis genes *Spo11* and *Hop1* are present in all major lineages of protists (Weedall and Hall, 2015), having been detected in Apicomplexa (Malik *et al.*, 2008), Euglenozoa (Malik *et al.*, 2008), Fornicata (Ramesh, Malik and Logsdon, 2005), Parabasalia (Malik *et al.*, 2008), and Amoebozoa (Ehrenkaufer *et al.*, 2013). On the other hand, the presence of meiosis genes does not necessarily prove that a species is sexual; orthologs may have more than one function or may adapt to perform new functions in asexual species and thus be maintained, even in the absence of sex (Schurko and Logsdon, 2008). Nevertheless, association of genes with a given function is very recurrent in molecular genetics based on neo-Darwinism, representing a very ineffective thinking strategy for understanding biological processes. It is worth remembering that *Drosophila melanogaster* does not have many meiosis genes, and this has not led developmental biologists to question the sexuality of arthropods (Malik *et al.*, 2008).

To support our hypothesis, it is important to search for an association between meiosis and the formation of different cells (anisogamy) in protists. A link would therefore be established between the first animal zygote and differentiation of oocytes and spermatozoa of the first metazoan embryo that would have been hermaphrodite. Protists have extensive biological experience in the formation of singular potentials that result in gametogenesis (sperm and oocyte phenotypes) and, particularly, differentiation of vegetative cells into gametes. Gametogenesis in *Chlamydomonas*, for example, is a phenomenon that allows the study of sexual differentiation in a single cell at the molecular level. Such differentiation is easily controlled by environmental signals, including lack of nitrogen and light (Beck and Haring, 1996). Some anisogamous species, such as *Chlamydomonas suboogama*, produce large immobile macrogametes and small mobile microgametes (a clear case of anisogamy) (Tschermak-Woess, 1959). Microgametes merge with macrogametes.

Some other anisogametic genera of flagellated protists include *Eudorina* and *Volvox*, which are either dioecious or monoecious. In monoecious species, each individual colony is either female or male, whereas, in dioecious species, a clone produces either female or male colonies (Tarín and Cano, 2000). Another example of anisogamy can be found in the Apicomplexa group. Infectious sporozoites invade host cells and then produce large and small gamonts. Large macrogamonts turn into macrogametes, and small gamonts undergo fissions, resulting in many small microgametes. Fertilization occurs by penetration of the microgamete into the macrogamete (Tarín and Cano, 2000). Among unicellular holozoans, *Salpingoeca rosetta* is the only species known to have a sexual life cycle, transitions between haploid and diploid states, and anisogamous mating, during which small flagellated cells fuse with larger flagellated cells (Levin and King, 2013).

Although there are currently no studies on the sexual reproduction of some groups of unicellular holozoans, such as Ichthyosporea, Filasterea, and Pluriformea, we believe that they have some cellular characteristics and that their life cycles are of extreme relevance in the context of our hypothesis. The amoeboid forms of *Corallochytrium limacisporum*, *Creolimax fragrantissima*, *Sphaeroforma arctica*, *Capsaspora owczarzaki*, *Ministeria vibrans*, and *Syssomonas multiformis* are of interest for the development of the hypothesis on the formation of the first embryo (Mendoza, Taylor and Ajello, 2002; Paps and Ruiz-Trillo, 2010; Sebé-Pedrós *et al.*, 2013; Suga and Ruiz-Trillo, 2015; Hehenberger *et al.*, 2017). Also of interest to our proposal are the flagellated forms of *Rhinosporidium seeberi*, *Dermocystidium* sp., *Dermocystidium salmonis*, *Dermocystidium percae*, *Sphaerothecum destruens*, *M. vibrans*, *Pigoraptor vietnamica*, and *Pigoraptor chileana* (Mendoza, Taylor and Ajello, 2002; Marshall *et al.*, 2008; Torruella *et al.*, 2012; Hehenberger *et al.*, 2017; Mylnikov *et al.*, 2019), as well as the flagellated amoeboid form of *S. multiformis* (Hehenberger *et al.*, 2017). The extremely complex life cycles of these groups are noteworthy, with emphasis on the colonial forms of *C. fragrantissima*, *S. arctica*, *P. vietnamica*, and *P. chileana* (Marshall *et al.*, 2008; Paps and Ruiz-Trillo, 2010; Hehenberger *et al.*, 2017) and the "syncytial" form of *C. fragrantissima* and *S. arctica* (Suga and Ruiz-Trillo, 2015) (syncytium-like structure may be a more appropriate term, given that it does not involve a fusion process), all of which demonstrate a successful attempt of unicellular organisms to remain together with a cellular structure that has numerous genes associated with Neoplasia (Domazet-Lošo and Tautz, 2010; Chen *et al.*, 2015; Grau-Bové *et al.*, 2017; Bartas *et al.*, 2019).

Fertilization is highly conserved in eukaryotes and animals (Cavalier-Smith, 1978; Carvalho-Santos *et al.*, 2011). We understand that flagella were co-opted to gamete fusion and therefore to the germline formation. Flagella are sensory structures in protists (Marshall and Nonaka, 2006; Singla and Reiter, 2006; Brunet and Arendt, 2016; Hilgendorf, Johnson and Jackson, 2016; Smith, Lake and Johnson, 2020). Some of their original characteristics were conserved in metazoans, such as chemotaxis (Miller and Brokaw, 1970; Ward *et al.*, 1985). For movement, *M. vibrans* does not use flagella but rather microvilli (Mylnikov *et al.*, 2019). As will be discussed in a future article, multiflagellate fusion might have been decisive in the evolution of metazoans. From the point of view of our hypothesis, the recruitment of cilia/flagella to NFM was crucial for animal evolution.

It is important to reflect on the reasons for excluding choanoflagellates as the original group of metazoans. Our line of reasoning is based on the contributions of Manuel Maldonado (Maldonado, 2005), who attributed numerous reductions and losses of this protist group and whose line of thought was confirmed in a recent phylogenetic study that placed choanoflagellates as the most distant group, outside the direct line of metazoans (Arroyo *et al.*, 2018). On the other hand, reassessment of molecular and histological evidence demonstrated that choanocytes are specialized cells that develop from non-collared ciliated cells during sponge embryogenesis, discarding choanocytes as primitive cells in animal evolution (Maldonado, 2005). Thus, the connection of choanoflagellates with metazoan emergence is a completely unrealistic idea.

We also cannot disregard the important contribution of Alexander Eresvkovski in expanding the knowledge about sponges (Ereskovsky, 2010; Ereskovsky, Renard and Borchiellini, 2013). As co-author of a recent publication, Eresvkovski described the differences between the flagellar apparatus (or kinetid) of sponges and that of choanoflagellates. According to the authors, analysis of these structures was inexplicably neglected (Pozdnyakov *et al.*, 2017). The kinetid of choanocytes contains more features considered plesiomorphic for opisthokonts than the kinetid of choanoflagellates. Therefore, the hypothesis that Porifera, and consequently all metazoans, originated directly from unicellular choanoflagellates does not seem plausible (Pozdnyakov et al., 2017).

Thus, according to our hypothesis and phylogenetic studies, the unicellular holozoans Ichthyosporea, Filasterea, and Pluriformea would be closer and in the direct line of metazoans (Suga *et al.*, 2013; Grau-Bové *et al.*, 2017; Hehenberger *et al.*, 2017). Considering these groups brings an innovative element that reinforces the idea of the link between fungi and animals (Maldonado, 2005). We are aware that yet undiscovered intermediate organisms could be in the direct line of metazoans; however, our hypothesis was developed on the basis of current data, and we await confirmation of meiosis associated with anisogamy in other groups of unicellular holozoans. In any case, we consider anisogamy as a condition for the biological prototype that would come to contribute to the establishment of the first animal embryo. Anisogamy is quite primitive among protists (Tschermak-Woess, 1959; Tarín and Cano, 2000) and was conserved in metazoans.

Finally, if we accept the premise of an NP at the heart of animal embryogenesis, we can explain the rapid evolution of forms of sex determination by contradicting one of the main myths of animal evolution (that it is slow and gradual) (Bachtrog *et al.*, 2014). NP consolidation might have had effects at the cellular level, including gene regulation, or might have provided for broad genomic recombination, thereby explaining why, sometimes, closely related species or populations of the same species have different modes of sex determination (Charlesworth, 1996). On the other hand, the NP at the core of animal embryos may help explain the mysterious preponderance of sexual reproduction between species (the õsex paradoxö), revealing that sexually reproducing organisms always proliferate faster (Chen and Wiens, 2021). As evidenced by the conditions for the formation of the first embryo, the association of multicellularity and sexual reproduction would make it possible to explain the rapid evolutionary radiation (Chen and Wiens, 2021). In our view, multicellularity and sexual reproduction can only explain the rapid formation of metazoans if strictly and intimately linked to the context of animal embryogenesis.

*The second condition: the germline formation*

Gametogenesis plays a unique role in gamete production and has a great impact on the heredity of embryogenesis, whose goal is the reconstruction of the embryo with each generation. Because it is a process of rebuilding the structural coherence of the embryo, it also plays a key role in evolution. Therefore, understanding the mechanisms of germ cell specification is a central challenge in developmental biology and evolution research.

During animal embryogenesis, the germline is segregated from somatic cells, but this dichotomy is not universal in animals. Recent findings in ctenophores (Fierro-Constaín *et al.*, 2017), poriferans, and cnidarians suggest that the germline appears to specify from a multipotent cell line during embryogenesis, whose fates include somatic cells and the germline (Juliano and Wessel, 2010). Thus, the boundary between the germline and the soma is fluid and may have a broader significance for development than initially predicted, being important, for instance, in active regeneration processes (Edgar, Mitchell and Martindale, 2021).

Another fundamental aspect is the exact moment of germline segregation. In animals, it seems to occur along a continuum, shortly after embryogenesis or at the very beginning of it, with the possibility of intermediate points during embryogenesis (Seervai and Wessel, 2013). In all cases, germline segregation requires that a population of cells be established in the embryo, whether multipotent or not. Given that germline segregation from a multipotent precursor occurs after embryogenesis in ctenophores, poriferans, cnidarians, and even in lophotrochozoans and echinoderms, it is inevitable to conclude that this is truly an ancestral mechanism (Agata *et al.*, 2006; Juliano, Swartz and Wessel, 2010).

Late segregation after embryogenesis is completely expected and necessary for the beginning of animal life, being one of the conceptual bases of our hypothesis. The first embryo and its germ cells must have first received the mechanical and physical stimuli of embryogenesis to trigger germline segregation. Therefore, the philosophical question should be, "What came first, the egg or the beroid-like ctenophore?" The answer is unequivocal; the egg came first, but with one caveat: only after it had been completely impregnated by its surroundings and physical trajectory inside the embryo. This first embryonic trajectory is fundamental: by walking, the path is made (Machado, 2021). This rule is absolute for the first embryo, which would come to establish a topological map of physics in the genetic material (Uhler and Shivashankar, 2017b, 2017a, 2018; Tsai and Crocker, 2022).

Some clues of this physical map indicate a primitive chromatin state after fertilization (Xia and Xie, 2020) that was highly organized, structured (Kaaij *et al.*, 2018), and conserved in evolution (Hug *et al.*, 2017; Kaaij *et al.*, 2018). Such a state seems to be important for the embryo to reach totipotency (Flyamer *et al.*, 2017). This organization occurs through topological associating domains (TADs) and chromosomal loops present in germ cells and animal zygotes (Flyamer *et al.*, 2017; Collombet *et al.*, 2020). Such a process is compatible with the physical impregnation of the embryo and its impact on germ cells. Another clue from the physical map is the extensive remodeling of chromatin after fertilization, which involves *de novo* labeling/trimethylation of histone H3 at lysine 9 (H3K9me3) by the G9a/GLP complex. This labeling facilitates subsequent establishment of a mature constitutive chromatin (Burton *et al.*, 2020). In other words, they seem to be "bookmark promoters for future compaction" (Xia and Xie, 2020), thereby "creating a less constrained epigenetic environment for subsequent zygotic genome activation" (C. Wang *et al.*, 2018). Once heterochromatin is established, H3K9me3 plays an important role in genome stability and cell differentiation fidelity (Nicetto *et al.*, 2019). We understand that this ability of H3K9me3 to anticipate what comes next in embryonic development (mechanical memory) is only possible because it recreates the physical map of the germline after fertilization. Finally, an important clue to the physical impact of embryogenesis lies in the cohesins that mediate the formation of chromosomal loops and TADs and are sensitive to mechanical force (Kim *et al.*, 2019). Polycomb repressive complex 2 (PRC2), which interacts physically and functionally with G9a/GLP (Mozzetta *et al.*, 2014), participates in a mechanosensory mechanism dependent on F-actin and the protein emerin (Le *et al.*, 2016)

There is clear evidence that TADs and, therefore, chromatin architecture are deeply interrelated with gene regulation (Dixon *et al.*, 2012; Nora *et al.*, 2012; Rao *et al.*, 2014; Fraser *et al.*, 2015; H. Wang *et al.*, 2018; Esposito *et al.*, 2020; Belyaeva *et al.*, 2022). The idea that the genome may be represented, in part, as a series of chromosomal blocks that can be opened or closed for transcription under specific conditions (Hurst, Pál and Lercher, 2004), among them physical ones, seems to be a new, useful, inspiring model. This is consistent with the notion that hematopoietic stem cells have the potential to generate non-hematopoietic tissues (Akashi *et al.*, 2003) and with labor division of yeast histone deacetylases in domains other than the chromatin (Robyr *et al.*, 2002). In mouse NIH 3T3 cells, actomyosin contractility regulates the spatial distribution of histone deacetylase-3 (Nikhil *et al.*, 2013). Therefore, physical forces can alter gene expression profiles and differentiation programs (Tajik *et al.*, 2016). Recent studies have shown that chromosomal configurations are altered in response to changes in nuclear mechanical properties following cues from the mechanical microenvironment (Farid *et al.*, 2019; Carthew *et al.*, 2021).

We speculate that mechanical changes (tension, substrate stiffness, or geometric constraints) are fundamental for the recruitment of what we call NFM (Figure 1). The module undergoes some co-optations and exaptations throughout the first animal embryogenesis, following the mechanical cues of embryogenesis. With the formation of the first embryo and its germline, there will be possibilities in the next generation to modify the trajectory built in the first embryo; thus, innovations of this topological map and NFM may arise at the beginning or middle of embryogenesis, producing a promising monster (Goldschmidt, 1982). Viruses (Breitbart *et al.*, 2015), symbionts (Margulis and Fester, 1991; Ohtsuka *et al.*, 2009), and physical or epigenetic disruptions can produce evolutionary innovations. But, for this to happen, physical phenomena must first have impregnated embryology.

One of the most fascinating current concepts is that of mechanical memory. The biophysical regulation of chromatin architecture produces stable remodeling and long-term changes of cellular behavior instigated by mechanical signals (Balestrini *et al.*, 2012; Heo *et al.*, 2015; Li *et al.*, 2017). Mesenchymal stem cells cultured for 3 weeks on a soft substrate produced a persistent neural differentiation resistant to other stimuli produced by cell differentiation factors (Engler *et al.*, 2006). Short-term mechanical memory is dependent on actomyosin contractility. Long-term increases in deformation that persistently affect chromatin condensation do not seem to depend on actin contractility but on calcium ion (Heo *et al.*, 2015). This phenomenon should be analyzed with care, as actin filaments gained reputation as bionanowires capable of conducting calcium waves (Hunley and Marucho, 2022). An important review study analyzed the different mechanisms responsible for storing mechanical memory in the cell nucleus (Dai, Heo and Mauck, 2020), including loops stabilized by CCCTC-binding factor (CTCF) or cohesins (Rao *et al.*, 2014; Kim *et al.*, 2019) and the epigenome (Le *et al.*, 2016), both of which are sensitive to mechanical force (Le *et al.*, 2016; Kim *et al.*, 2019). *In vivo* models of mechanical memory have been little explored, being used only in cancer research (Paszek *et al.*, 2005; Northey, Przybyla and Weaver, 2017). In this context of cellular memory, the order of NFM expression must apparently be determined by the 3D structure of the nucleus (H. Wang *et al.*, 2018), which in turn is influenced by the 3D structure of the embryo.

In relation to germ cells, special emphasis should be given to ctenophores of the order Beroida. The recognizable germ cells of ctenophores are closely associated with the endoderm of the meridional canals and always underlie the comb rows (Pianka, 1974). In beroids, gonadal tissues either form simple tracts in proper meridional canals, extend into the lateral branches that are characteristic of this order, or develop into separate sexual diverticula external to lateral branches. Thus, the presence of a multipotent cell line and a fluid relationship between soma and germline (Edgar, Mitchell and Martindale, 2021) may be explained by organization or physical construction of the embryo. Germ cells (established within the gonads) pass on or receive directly or indirectly most of the physical impacts and tension of the embryo (e.g., epiboly, invagination, and EMT, among others, as will be discussed below) (Figure 1). Luke Parry's analysis provides insight about a sensory network in the first beroid-like ctenophores (Parry *et al.*, 2021). The main highlights regard the nerve tracts of comb rows and polar fields, showing a much more complex sensory and nervous organization than that of living ctenophores (Parry *et al.*, 2021). This type of network can reflect on the physical, mechanical (tensional actin networks), and electrical (nerve impulses) organization of the embryo. The implications of such a network may be depicted in the ability to regenerate. Surprisingly, *Beroe ovata* lost nerve endings in comb rows as a byproduct of evolution (Parry *et al.*, 2021), thereby losing the ability to regenerate (Edgar, Mitchell and Martindale, 2021). Mechanical forces may be morphogenic and organize how cells regenerate. There is strong evidence that regeneration capacities are related to the apical (sensory) organ, when present (Martindale, 2016). Mark Martindale suggested that nervous system involvement may be crucial for ctenophore regeneration and proposed experimental manipulation to confirm or rule out neural involvement (Edgar, Mitchell and Martindale, 2021).

Although the so-called germline multipotency program is conserved in ctenophores (Fierro-Constaín *et al.*, 2017), which involve, among other genes, *vasa*, *piwi*, and *nanos*, these genes show little or no expression in germ cells or gametogenic regions (Reitzel, Pang and Martindale, 2016). Restricted expression in the apical organ and tentacles of *Mnemiopsis leidyi*, which are areas of high cell proliferation, suggests that these genes could be involved in the specification and maintenance of stem cells (Reitzel, Pang and Martindale, 2016). Similarly, members of the gene family *Dmrt*, essential components of sex determination and differentiation in bilateral animals, were not expressed in adult gametogenic regions.

What becomes obvious from these analyses is that *vasa*, *nanos*, and *piwi* genes do not participate in the establishment or maintenance of the germline of animals. Ctenophores, for example, produce haploid gametes and reproduce sexually; thus, their germline has the same function as that of bilaterian animals. That is, it is not this set of genes that determines germline development and maintenance but rather the appropriate cellular context in which the genes function. The appropriate context for *vasa*, *nanos*, and *piwi* was established a little later in animal evolution and thus co-opted by embryonic germ cells for more specialized functions (Juliano and Wessel, 2010). The genes responsible for maintaining and establishing the germline in ctenophores remain unknown, but evolution has left some traces and clues of their origin.

Some evidence on the establishment of the germline in basal metazoans emerged with the identification of a highly conserved system in invertebrates and vertebrates related to the protein Mos and its function in meiosis (Amiel *et al.*, 2009). Mos, curiously a protooncogene, would prevent meiotic/mitotic conversion (meiotic to mitotic conversion) after meiosis I, directing cells to meiosis II and ensuring ploidy reduction (Tachibana *et al.*, 2000), thereby preventing undesirable DNA replication or parthenogenetic activation prior to fertilization (Furuno *et al.*, 1994; Tachibana *et al.*, 2000; Amiel *et al.*, 2009). Mos participates in induction of oocyte maturation (Freeman *et al.*, 1990) and migration of the meiotic spindle (Verlhac *et al.*, 2000). These functions are crucial for two very unique moments of embryogenesis: (i) fertilization (ensuring ploidy reduction before the onset of embryogenesis) and (ii) separation of the germline from the somatic lineage, when meiosis begins (soma-to-germline transition). This leads us to speculate on the role of Mos, or a similar protein, in establishing the germline in the first embryo. Unfortunately, Mos has not yet been studied in the cell cycle or meiosis of ctenophores but was characterized in a genomic analysis of *Pleurobrachia pileus* (Amiel *et al.*, 2009). Cytoplasmic tyrosine kinases (among them Mos/MAPK) seem to have been established before divergence of Metazoa (Suga *et al.*, 2012).

The protein Mos serves to remind us that meiosis evolved from mitosis and that it could also alleviate the polyploidy instability (Cleveland, 1947) (Figure 2e). In fact, endomitosis is a form of meiosis but without karyogamy (Raikov, 1995). In the flagellated protists *Trichonympha* and *Barbulanympha* (Cleveland, 1947) and the radiolarian *Aulacantha scolymantha* (Grell and Ruthmann, 1964), there is evidence of a prototype of chromosomal pairing in endomitosis (Figure 2d). As reported by Lemuel Cleveland, polyploidy in *Barbulanympha* is invariably reduced by meiosis (Figure 2e), and it is neither preceded nor followed by any type of syngamy (Cleveland, 1947). It is also recognized that, for endomitosis to occur, chromosome condensation and telomere duplication are important, which clearly distinguishes it from polyteny (Erenpreisa, Kalejs and Cragg, 2005). It has already been hypothesized that Mos would be part of the molecular basis for somatic reduction and a return to the mitotic cycle of endopolyploid tumor cells (Erenpreisa, Kalejs and Cragg, 2005). This proposal by Jekaterina Erenpreisa and Mark Cragg is consistent with results in two-cell embryos from *Clytia hemisphaerica*, in which "lower concentrations of Mos RNA caused cleavage arrest with multiple nuclei or spindles within a common cytoplasm," demonstrating the occurrence of Mos-mediated endomitosis in Cnidaria (Amiel *et al.*, 2009), consistent with coenocytic division carried out by our closest relatives, ichthyosporeans (Ros-Rocher *et al.*, 2021) (Figure 2c). Also, the blockade of Mos in oocytes produces a "parthenogenetic development with complete cleavages" but the presence of "multiple asters and spindle poles during mitotic cycling," further corroborating the presence of endomitosis in *Asterina pectinifera* (starfish) (Tachibana *et al.*, 2000). The multiple asters and spindle poles are also present in the megakaryocyte endomitosis and possibly arise due to alterations in the regulation of mitotic exit (Vitrat *et al.*, 1998), known to be controlled by Mos. In our NFM model, we speculate the control of three types of division (mitosis, meiosis, and endomitosis) and the recruitment of Mos (or a similar protein) would have been essential for germline emergence and incorporation of polyploidy in animal embryogenesis. As will be discussed in another article, the control of these types of cell division, including the two types of meiosis (meiosis and endomitosis), was fundamental to the success of the multicellular embryo.

According to our proposal, for the embryo to achieve reconstruction and structural coherence in following generations, it must necessarily incorporate NP (through NFM) into the cellular context of germ cell formation. In other words, the germline is closely linked to NP, because it is at the heart of animal formation. Therefore, germline and NP are impossible to separate in the context of metazoans. We may even predict that germline gene expression will manifest in the context of somatic cancers and that germline cancers may tend to produce arrangements that can be "interpreted as attempts at organization" (Pierce, 1967).

Consistent with our proposal, several studies on *D. melanogaster* (Janic *et al.*, 2010), mice (Ma *et al.*, 2012), and humans (Koslowski *et al.*, 2004; Rosa *et al.*, 2012; Rousseaux *et al.*, 2013) revealed that tumors acquire a state very similar to that of the germline, indicating that tumorigenesis in metazoans involves a soma-to-germline transition, possibly contributing to the acquisition of neoplastic characteristics (Wang *et al.*, 2011; Feichtinger, Larcombe and McFarlane, 2014). It is important to remember that the boundary between soma and germline is fluid (Edgar, Mitchell and Martindale, 2021) and associated with a neoplastic force for embryo construction. It was demonstrated that germline gene expression has the potential to be oncogenic in *D. melanogaster* (Janic *et al.*, 2010) and is associated with more clinically aggressive tumors in humans (Rousseaux *et al.*, 2013). Inactivation of some germline genes results in suppression of tumorigenesis, indicating that they play an essential role in tumor development (Janic *et al.*, 2010). Finally, the soma↔germline transition was also observed in *Caenorhabditis elegans*, suggesting a conserved functional relationship between tumorigenesis and germline gene expression in metazoans (Unhavaithaya *et al.*, 2002; Wang *et al.*, 2005). Such relationships are sensible in the context of our proposal of NP as an evolutionary engine.

Also consistent with our hypothesis is that, in germline cancers, two different forces oppose each other: a force of embryonic organization and differentiation contained in the germ cell structure (neoplastic process) and a disruptive force of cancer (disease). In this regard, the works of G. Barry Pierce are very inspiring. The author demonstrated that, in a small area of testicular teratocarcinoma, the mesenchyme and endoderm are often organized in patterns that resemble early stages of embryogenesis, having been called embryoid bodies (Dixon and Moore, 1953; Pierce, Dixon Jr and Verney, 1959; Pierce, 1967). In the works of Winston Evans, some embryoids "closely mimic the form of normal early human embryos" (Evans, 1957). A conceptual basis for the developmental model of cancer was well documented in Pierce's works; thus, the dogmas that neoplastic forces are stable and irreversible were broken. It was hypothesized that cancer is controlled by intrinsic mechanisms of the embryo (embryonic induction) (Pierce, 1967, 1983; Arechaga, 2003). In the context of this article, we extend the concept to propose that the embryo and cancer are the result of an equilibrium achieved within the NFM during embryogenesis and at the beginning of the evolution of metazoans.

Embryogenesis records will be restricted to a cellular set that incorporates NFM and can later reproduce and restart in a coherent and systematic way the construction of a new animal, with the same structural form as the reproduced one. This is a central aspect that clearly separates any multicellular organization from a true animal embryo.

**Concluding remarks and perspectives**

In this first article, we established that the emergence of multicellularity only makes sense if we explain the formation of the first embryo. We also begin to show the theoretical basis that supports the close link between embryogenesis and NP. One of the central aspects of this paper is to begin to link the impact of Physics and environmental conditions responsible for the formation of the first embryo. Post-fertilization analyses point to a highly organized genomic structure with TADs and loops (Flyamer *et al.*, 2017; Hug *et al.*, 2017; Kaaij *et al.*, 2018; Collombet *et al.*, 2020; Nakamura *et al.*, 2021), but without compartments (Chen *et al.*, 2019; Collombet *et al.*, 2020), an organization that is concordant with an NFM present in the germline and active early in embryogenesis. Some of the most interesting aspects clarified in the last decade were that the 3D organization of the cell nucleus affects gene regulation (Zuin *et al.*, 2014; Varun *et al.*, 2015; Zheng and Xie, 2019; Rhodes *et al.*, 2020) and that the formation of loops and TADs by cohesins is influenced by physical forces (Kim *et al.*, 2019). Application of a force of 800 pN on cohesins I prevented the formation of chromosomal loops. G9a/GLP and PRC2 are the largest epigenetic silencing complexes, which methylate histone H3 at lysines 9 and 27, respectively (H3K9 and H3K27) (Mozzetta *et al.*, 2014), forming part of a mechanosensory mechanism (Le *et al.*, 2016). That is, it is impossible to disconnect the nuclear structure from the physical context, ECM, and physical interactions of the tissue.

The main challenges are to study chromosomal rearrangement on a large scale during embryogenesis, considering physical phenomena and how the nucleus is interconnected to a network of cellular and tissue cytoskeleton. The removal of pro-nuclei from a zygote, by suction with a micropipette or dissociation of blastomeres from an embryo, may alter the 3D nuclear morphology. Therefore, models must be adjusted to understand how chromosomal organization is influenced by mechanical and electrical phenomena considering the entire embryo. Alternatives to CRISPR genome organization (H. Wang *et al.*, 2018) and mathematical modeling at the embryonic level (Belyaeva *et al.*, 2022) begin to take on importance for the analysis of how biophysics modulates animal architecture in ontogeny and phylogeny.

**Acknowledgments**

We thank Dr. Jose Bastos, pathologist and oncologist, for his permanent contribution to our cancer research projects. The authors offer apologies to all researchers who were not mentioned in the article, given the need to establish priorities in the article's construction.

**Conflict of Interest Statement**

The authors declare that the research was conducted in the absence of any commercial or financial relationships that could be construed as a potential conflict of interest.

embryos', *Genome Research*, 31(6), pp. 968–980. doi: 10.1101/gr.269951.120.

Nakatsuji, N., Smolira, M. A. and Wylie, C. C. (1985) 'Fibronectin visualized by scanning electron microscopy immunocytochemistry on the substratum for cell migration in Xenopus laevis gastrulae', *Developmental Biology*, 107(1), pp. 264–268. doi: 10.1016/0012-1606(85)90395-1.

Nanjundiah, V. (2016) 'Cellular Slime Mold Development as a Paradigm for the Transition from Unicellular to Multicellular Life', in Niklas, K. J. and Newman, S. A. (eds) *Multicellularity: origins and evolution*. Cambridge, Massachusetts: MIT Press, pp. 105–130.

Newman, S. A., Forgacs, G. and Müller, G. B. (2006) 'Before programs: The physical origination of multicellular forms', *The International Journal of Developmental Biology*, 50(2–3), pp. 289–299. doi: 10.1387/ijdb.052049sn.

Nicetto, D. *et al.* (2019) 'H3K9me3-heterochromatin loss at protein-coding genes enables developmental lineage specification', *Science (New York, N.Y.)*. 2019/01/03, 363(6424), pp. 294–297. doi: 10.1126/science.aau0583.

Nikhil, J. *et al.* (2013) 'Cell geometric constraints induce modular gene-expression patterns via redistribution of HDAC3 regulated by actomyosin contractility', *Proceedings of the National Academy of Sciences*, 110(28), pp. 11349–11354. doi: 10.1073/pnas.1300801110.

Nora, E. P. *et al.* (2012) 'Spatial partitioning of the regulatory landscape of the X-inactivation centre', *Nature*, 485(7398), pp. 381–385. doi: 10.1038/nature11049.

Northey, J. J., Przybyla, L. and Weaver, V. M. (2017) 'Tissue Force Programs Cell Fate and Tumor Aggression', *Cancer Discovery*, 7(11), pp. 1224–1237. doi: 10.1158/2159-8290.CD-16-0733.

Ohtsuka, S. *et al.* (2009) 'Symbionts of marine medusae and ctenophores', *Plankton and Benthos Research*, 4(1), pp. 1–13. doi: 10.3800/pbr.4.1.

Paps, J. and Ruiz-Trillo, I. (2010) 'Animals and Their Unicellular Ancestors', in
43

FIGURES

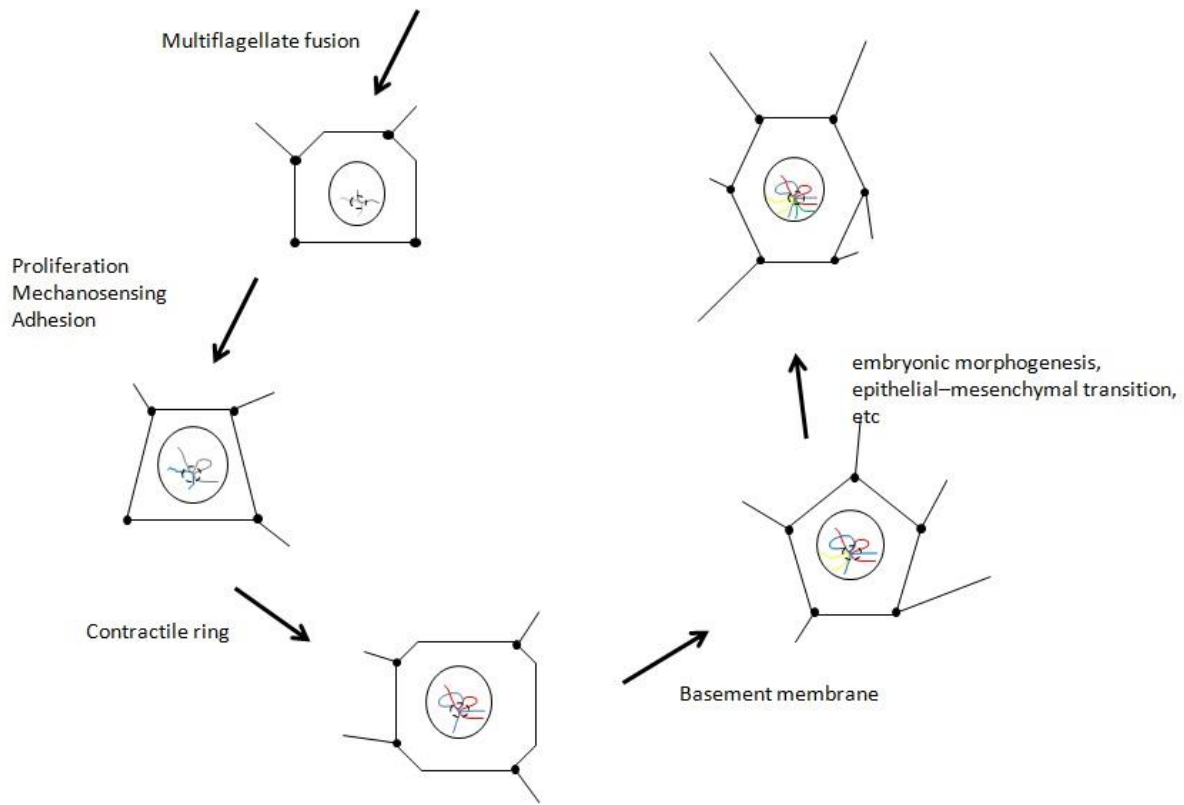

Figure 1. General scheme of the neoplastic functional module. Chromosomal domains were associated topologically by the physical impact of embryo construction. The projected points

and lines of the geometric figures represent the impact of embryonic morphogenesis on the cell nucleus. The cells depicted in the figure are multipotent cells that receive most of the biophysical impact promoting embryo patterning. Neoplasia is the driving force of embryo formation. The disease cancer is imbued in embryo construction and masked by its organization. The last cell of the scheme is one of the two cells of the hermaphroditic germ line of the first embryo, which harbors the neoplastic functional module and mechanical memory, elements that contribute to the reconstruction of the process in the following generation. Emergence of animal phylogeny, cancer, and the first embryo is implicit in the events represented in this diagram.

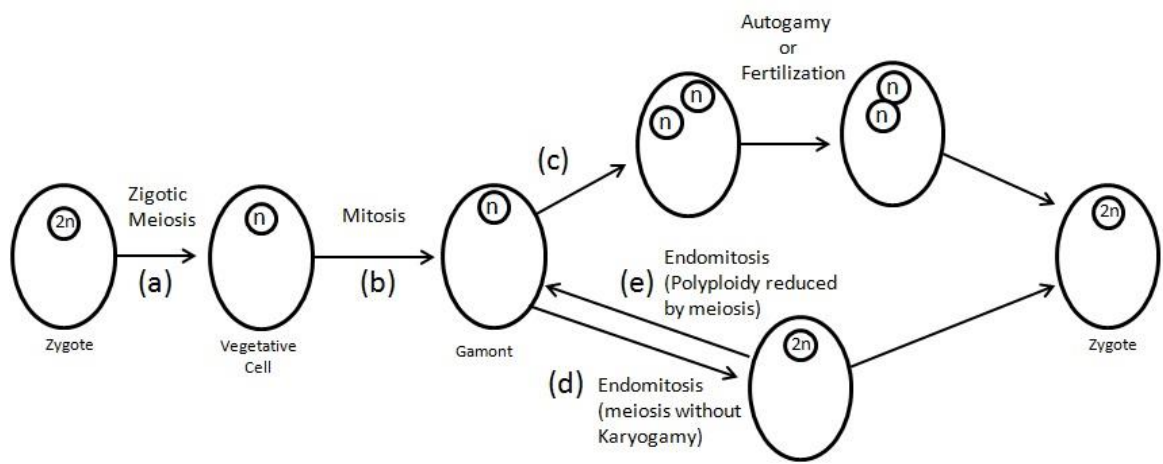

Figure 2. Zygotic meiosis and its modifications in sexual reproduction. (a) The zygote divides by meiosis to produce a vegetative cell. (b) Gamonts are produced by mitosis of vegetative cells. (c) A coenocytic-like division can produce two haploid nuclei that, by autogamy, produce a zygote. In this case, the modification lies in the omission of cell division (also known as endomitosis). Autogamy and fertilization can produce a zygote. (d) A further simplification (which can scarcely still be considered "normal") consists in the total omission of nuclear division. What remains is the reduplication of chromosomes without spindle formation, characterizing a case of endomitosis. Endomitosis allows the formation of a diploid cell. (e) Through (reverse) endomitosis, it is also possible to reduce polyploidy.

Figure adapted from Grell (1973).